\documentclass[12pt,preprint]{aastex}

\begin{document}

\title{The Sloan Bright Arcs Survey: Four Strongly Lensed Galaxies 
with Redshift $>$ 2 }

\author{H.~Thomas~Diehl\altaffilmark{1}, Sahar~S.~Allam\altaffilmark{1},
James~Annis\altaffilmark{1}, Elizabeth~J.~Buckley-Geer\altaffilmark{1}, 
Joshua~A.~Frieman\altaffilmark{1,2},
Donna~Kubik\altaffilmark{1}, Jeffrey~M.~Kubo\altaffilmark{1},
Huan~Lin\altaffilmark{1}, Douglas~Tucker\altaffilmark{1}, 
Anderson~West\altaffilmark{1,3}}

\altaffiltext{1}{Fermi National Accelerator Laboratory, P.O. Box 500,
    Batavia, IL 60510}
\altaffiltext{2} {Department of Astronomy and Astrophysics, University
of Chicago, 5640 South Ellis Avenue, Chicago, IL 60637}
\altaffiltext{3}{The Illinois Math and Science Academy, Aurora, IL 60506}

\begin{abstract}
We report the discovery of four very bright, strongly-lensed
galaxies found via systematic searches for arcs  
in Sloan Digital Sky Survey Data Release 5 and 6.
These were followed-up with spectroscopy and imaging data from 
 the Astrophysical Research Consortium 3.5m
telescope at  Apache Point Observatory and found to have redshift $z>2.0$.
With isophotal magnitudes $r = 19.2 - 20.4$ and
$3\arcsec$-diameter magnitudes $r = 20.0 - 20.6$, these systems are
    some of the brightest and highest surface brightness 
    lensed galaxies known in this redshift range.
In addition to the magnitudes and redshifts, we present estimates of the 
Einstein radii, which range from $5.0 \arcsec$ to $12.7 \arcsec$, and use those
to derive the enclosed masses of the lensing galaxies.
\end{abstract}


\section{Introduction}
Strong gravitational lenses provide an opportunity for studying 
properties of distant galaxies. Because surface brightness is unchanged
during lensing, magnification of the image provides amplification of
the image flux and allows studies of details that would 
otherwise be too faint for detailed ground-based investigation.  
Models of the lens systems also provide information on the
mass distribution, including dark matter, of the lenses themselves.

Recent wide area optical surveys have enabled identification of 
new samples of strong lens systems. The Sloan Digital Sky Survey 
(SDSS)~\cite{york} spectroscopic data has been 
used~\cite{slacsfull,ols-lens} to identify $\sim 140$ strong lensing 
candidates in systems where multiple spectra appear in a 
single $3\arcsec$-wide spectral fiber. The SDSS imaging data provides the 
opportunity to identify a complementary set of strong lens systems that 
have a larger lens-arc separation $(> 3.0\arcsec)$.
An imaging search of 240 rich clusters~\cite{Hennawi} yielded
16 strong lens systems with large ($> 10\arcsec$) radius, where the lensing 
interpretation is based on morphology, and 21 additional systems for 
which the lensing interpretation is less certain.  Several more large arcs 
in the SDSS data have been described in recent papers~\cite{belokurov,wen}.
The CFHTLS field  and COSMOS field have yielded 40~\cite{Cabanac}
and 70~\cite{faure,jackson} strong lens candidate systems.

Relatively few arc systems have been reported where the lensed
galaxy is known to have redshift $z > 2$.  Some systems with high 
surface brightness have been reported.
The brightest Lyman Break Galaxy (LBG) currently 
known is SDSS J002240.91+143110.4 ``The 8 o'clock arc''~\cite{8oclock} 
with a  redshift $z=2.73$ and magnitude $r=19.2$.
Other bright systems include SDSS J120602.09+514229.5 
``The Clone''~\cite{clone} with redshift $z=2.0$ and magnitude $r=19.8$, 
MS 1512-cB58~\cite{cB58} with redshift $z=2.73$ and 
magnitude $r=20.4$, and the system~\cite{z307} 
lensed by MACS J2135.2-0102 with redshift $z=3.07$.  

Motivated by our discovery of ``The 8 o'clock arc'',
we initiated a program, the Sloan Bright Arcs Survey,
 to identify and confirm lensed high-redshift 
galaxies. We recently reported~\cite{kubo} the discovery of six 
spectroscopically confirmed systems with redshifts $z = 0.4$ to $1.4$.
In this paper we report the discovery and confirmation of four more
systems with redshift $z>2$.

This paper is organized as follows. 
In $\S 2$ we describe the strong-lens arc search samples and candidate 
identification procedure. In $\S 3$ we describe our follow-up imaging 
and spectroscopy procedure. In $\S 4$ we describe the results 
from the follow-up and use those results to calculate 
properties of the lensed galaxies. Finally, in $\S 5$, we describe further 
follow-up measurements underway and summarize.

\section{Arc Search Samples}
Between 2000 and 2008, the Sloan Digital Sky Survey imaged more 
than 8,400 square degrees of the extragalactic sky in $u$, $g$, $r$, $i$, and 
$z$-band filters and obtained spectra of 930,000 galaxies.  
We searched~\cite{Kubik} for strong-gravitational lens galaxies in the
SDSS Data Release 5 catalog~\cite{jen} around luminous red galaxies
(LRG's) \cite{eisenstein} and brightest cluster 
galaxies (BCG's)~\cite{Hansen}. 
These were sorted according to the number, $N$, of 
additional galaxies located within $10 \arcsec$ that have a 
color $g$-$r$ $< 1$ and $r$-$i$ $< 1$. 
The list of candidates was provided to a web-based display tool that allows 
one to extract small false-color images using the SDSS Catalog Archive 
Server (CAS), based on a user-defined list of positions. 
We inspected  the images to find ones where there was a spatially
continuous arc or ring, or a nearly spatially continuous set of knots, 
(the source candidate) that is centered or nearly centered on a galaxy 
or galaxy cluster (the lens candidate). We required that the source
candidate not have a distinct color-change over its length (to be 
consistent with a single lensed source interpretation) and that the
source candidate not be continuously-connected to the central object by 
spiral arms, bars, or non-structural lumination. 
This was to eliminate galaxies that are providing 
both the central structure and the arc, ring, or knot.  
Independent inspections were performed by four different people
and any candidate that was noted by two or more scanners was marked for 
further study. There were 14 such candidates.

In addition, a second list of candidates resulted from an independent
inspection of an SDSS DR6 catalog of merging galaxies. In this catalog, 
a merging galaxy pair
is defined~\cite{ssamerge} as two galaxies in the magnitude range 
$16.0 < g < 21.0$ separated by less than the sum of their respective 
Petrosian radii~\cite{stoto}. There are many
systems with morphology and color similar to our other strong lens 
candidates. 

\section{Follow-up Imaging and Spectroscopy}
We followed up the 14 candidates from the first list and a subset 
of the candidates from the second list using the Astrophysical Research 
Consortium (ARC)  3.5m telescope at the
Apache Point Observatory. Visible-wavelength imaging was performed using 
the SPICAM CCD imager, which has a field-of-view of 4.78 arcminutes square
and a plate scale of $0.28\arcsec$ per pixel. We observed each candidate 
system in the three SDSS filters $gri$, taking in each filter typically 
three 300-second exposures, dithered with $15 \arcsec$ offsets.  
The images were bias subtracted and flatfielded using standard IRAF tasks
and were then coadded using the SWarp package~(Bertin 2006).
Astrometric calibration was carried out using matches to objects in the SDSS.
Likewise, photometric calibration was done by matching to stars in the  SDSS.
Specifically, we used GALFIT~\cite{galfit} to fit Moffat profiles 
to stars in the images, and set the photometric zeropoint of an image by
taking the median offset between the Moffat profile total magnitudes 
and the SDSS model magnitudes for these stars.  Additional details of our 
reduction procedure may be found in reference~\cite{clone}.

We report magnitudes in $3\arcsec$-diameter circular apertures for
our arc systems (except one instance as noted), 
where the photometry was done using 
SExtractor~\cite{sextractor} in dual-image mode, using the $r$-band
image for each system to define object detection.  Note that due to
imperfect SExtractor deblending of the contaminating light of the lensing 
galaxies from that of the arcs, the magnitudes we quote for the arc 
components may be systematically too bright by a few tenths of a magnitude,
especially in the $i$ and $r$ images where the red lensing galaxies
are more prominent.  We will defer to future analyses more careful
photometric modeling of these systems \citep[i.e., using GALFIT to subtract
the lensing galaxy light, as in][]{clone} as a prelude to 
detailed lens modeling, but in the present discovery paper simple
aperture magnitudes will suffice as reasonable first approximations
to the photometry of the arc components. 
We also report SExtractor isophotal magnitudes for our arcs,
where the limiting isophotes are the SExtractor
$r$-band detection thresholds (1.5 times the image RMS). 
Unless specifically noted, all magnitudes reported are 
those from $3\arcsec$-diameter apertures.

Spectroscopy was obtained 
using the Dual-Imaging Spectograph (DIS III), a medium-dispersion
spectrograph with separate red/blue grating setups. The spectral range
is 3000 -- $9600$ \AA \ with a resolution of $1.83$ \AA \ per pixel in
the blue and $2.31$ \AA \ per pixel in the red. Bright arcs and central 
galaxies were targeted, typically with three 900 second exposures 
using a $1.5\arcsec$ or $2.0\arcsec$ wide slit.  
The spectra were reduced and extracted
using standard IRAF routines and were flux calibrated using observations
of the spectrophotometric standard G191-B2B.  Redshift measurements were
done using the IRAF external task {\sc xcsao}, 
by cross correlation against
the Lyman break galaxy (LBG) template of Shapley et al. (2003).
The cross correlation was limited to the observed wavelength range 
3500 -- 5500~\AA \ (except for one system, as noted below), where all 
the relevant LBG spectroscopic features were located for our systems.  
Note we calibrated all our spectra to air wavelengths, so that the 
LBG template had to be transformed from vacuum to air wavelengths before 
being used for cross correlation.
Dates for useful follow-up observations are listed
in Tables~\ref{tab-obs} and \ref{tab-spe}.

\section{Lensing Systems}

\begin{figure}
\plottwo{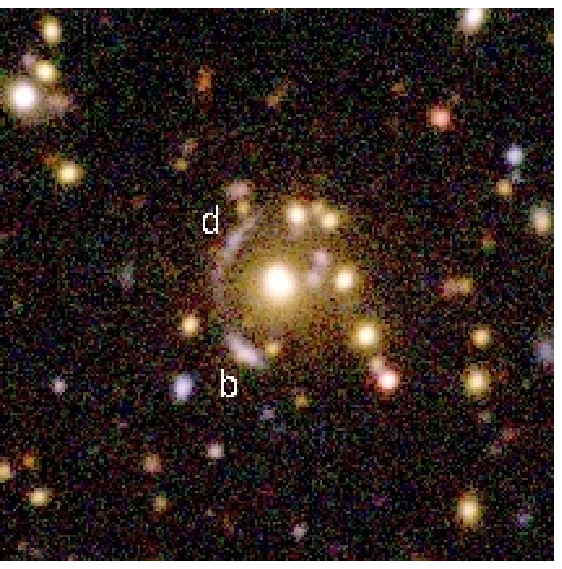}{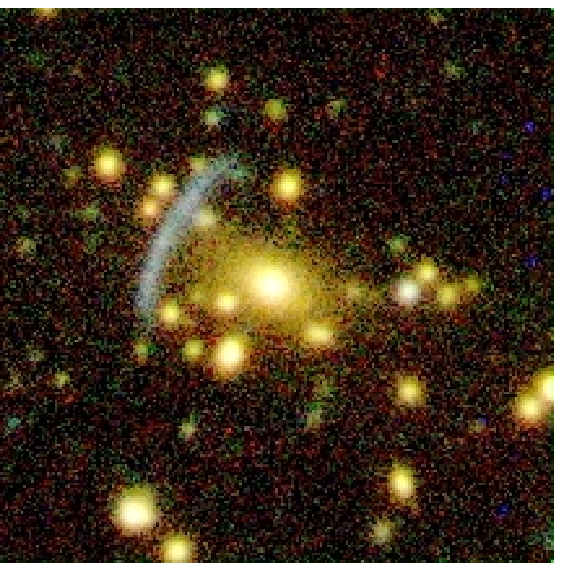}
\caption{Images of SDSS J090122.37+181432.3 (left)
and SDSS J134332.85+415503.4 (right).  The images are $1\arcmin$ x $1\arcmin$.
North is up. East is to the left. The knots are labeled to
distinguish between spectroscopic targets.
\label{fig-1}}
\end{figure}

\begin{figure}
\plottwo{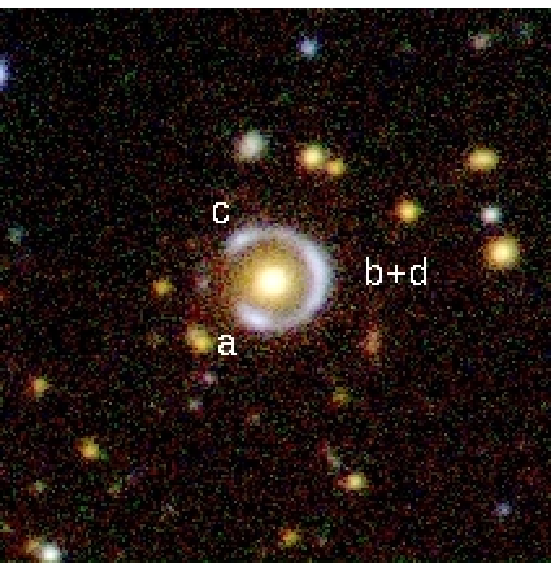}{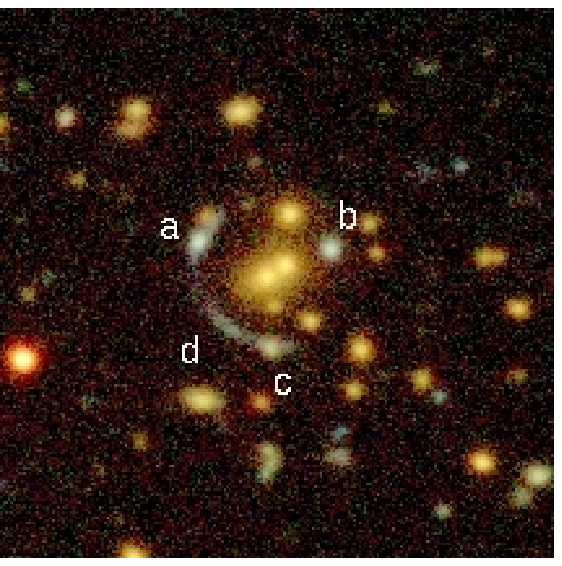}
\caption{Images of SDSS J114833.14+193003.2 (left) and   
SDSS J090002.79+223403.6 (right).  The images are  $1\arcmin$ x $1\arcmin$.  
North is up. East is to the left. 
 The knots are labeled to
distinguish between spectroscopic targets.
\label{fig-2}}
\end{figure}

\begin{figure}
\plotone{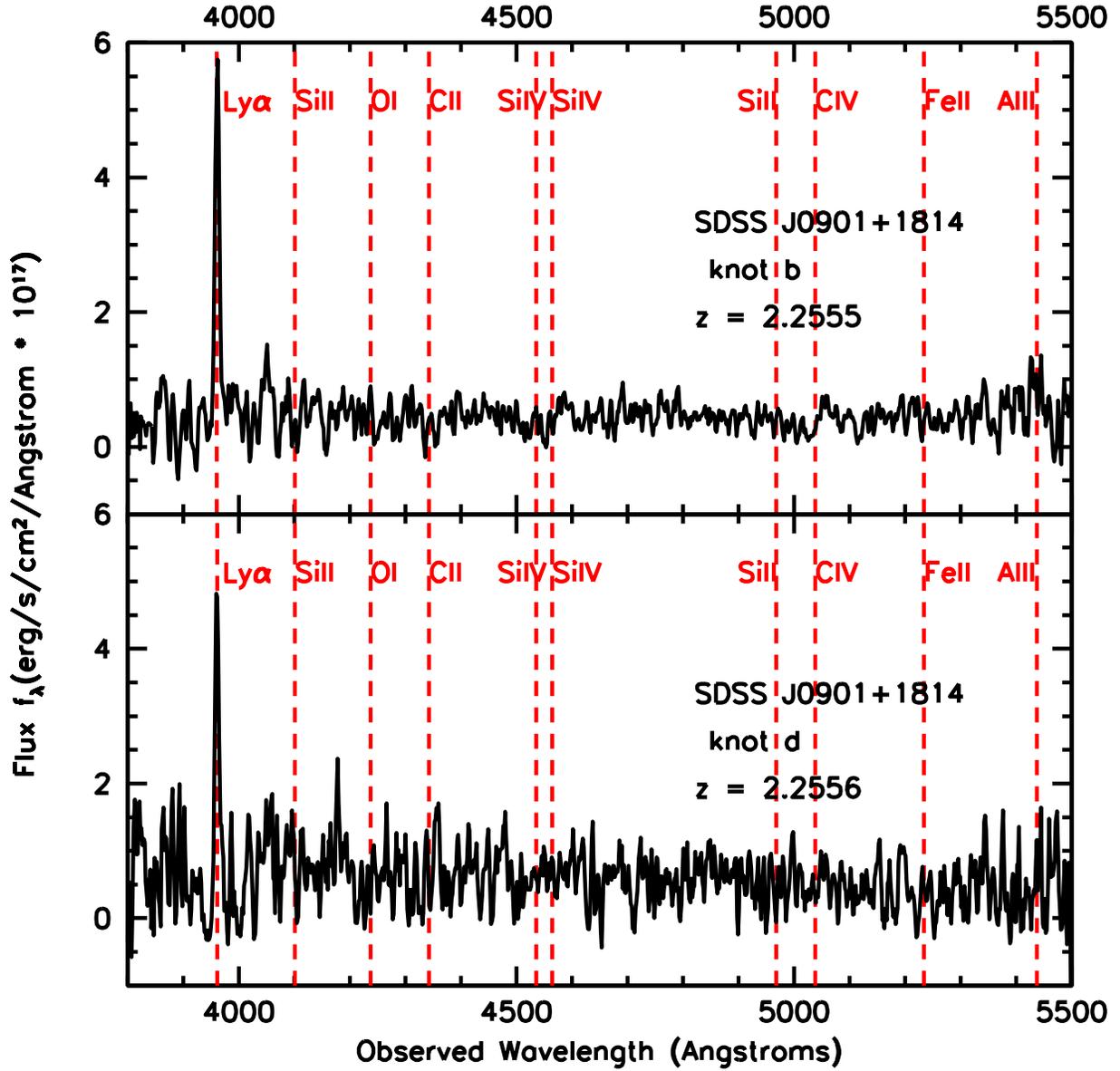}
\caption{Spectra for SDSS J090122.37+181432.3.
The upper panel is the spectrum for knot b as shown in Fig.~1 (left).
The lower panel is the spectrum for knot d.
\label{fig-3}}
\end{figure}

\begin{figure}
\plotone{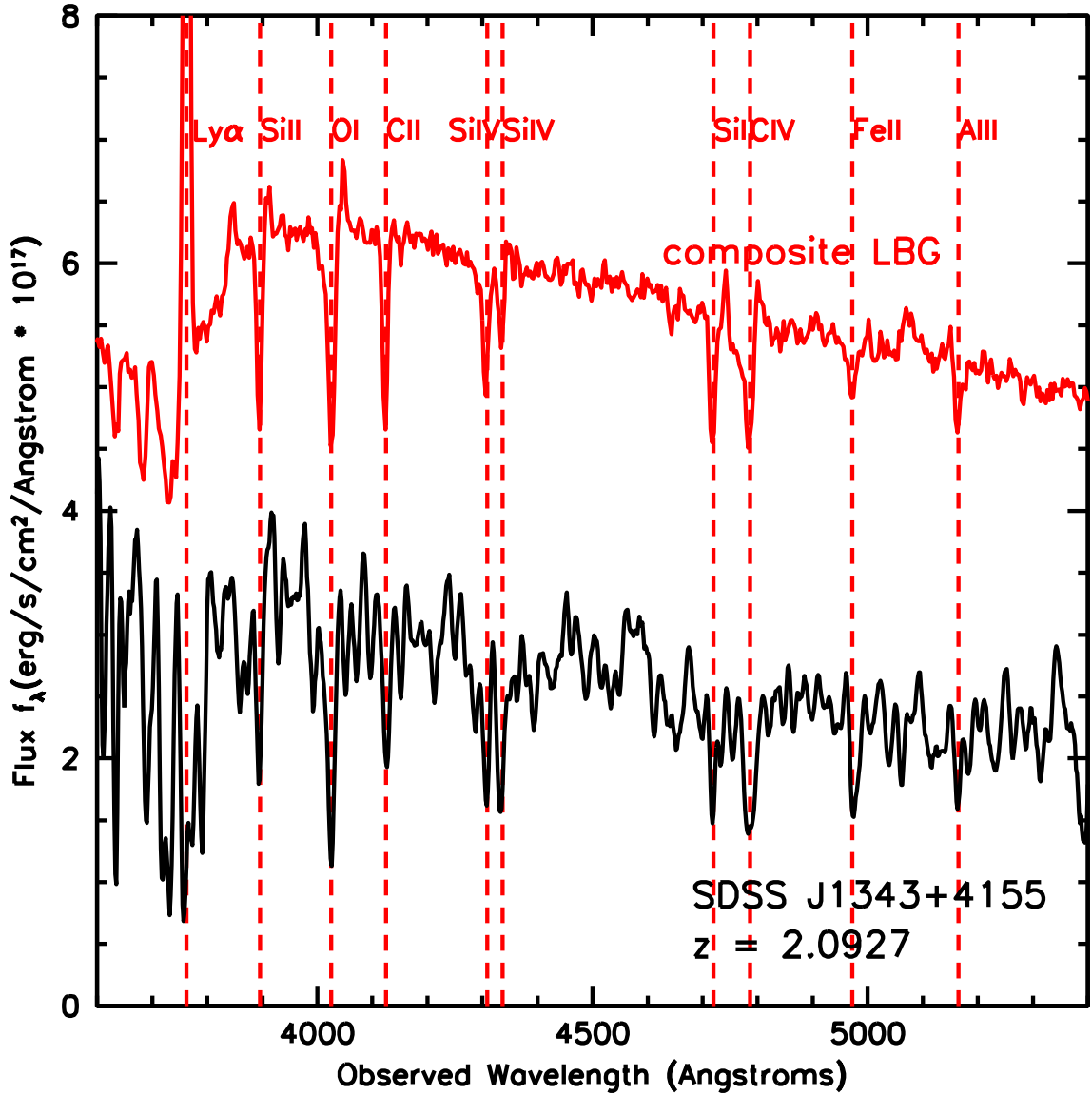}
\caption{Spectrum for SDSS J134332.85+415503.4. 
  The upper spectrum is the
composite LBG spectrum of Shapley et al. (2003). The lower spectrum is
the arc. We aligned the slit along the  arc.
\label{fig-4}}
\end{figure}

\begin{figure}
\plotone{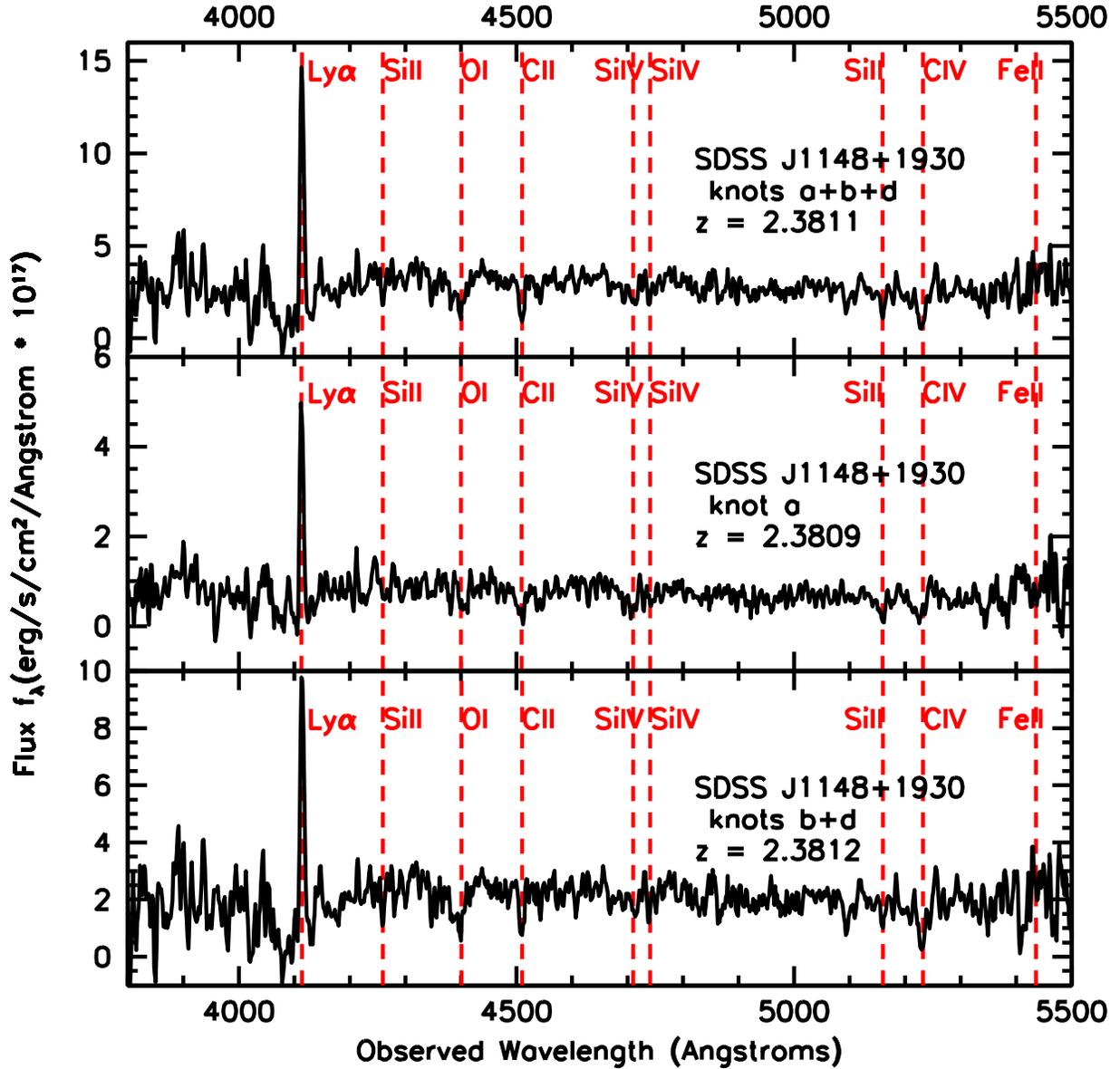}
\caption{Spectra for SDSS J114833.14+193003.2.
In the upper panel we show the sum of the spectra for knots a, b, and d
as shown in Fig. 2 (left). In the middle panel the slit was oriented
so as to cover only knot a. In the lower panel the slit was oriented
over knots b and d.
\label{fig-5}}
\end{figure}

\begin{figure}
\plotone{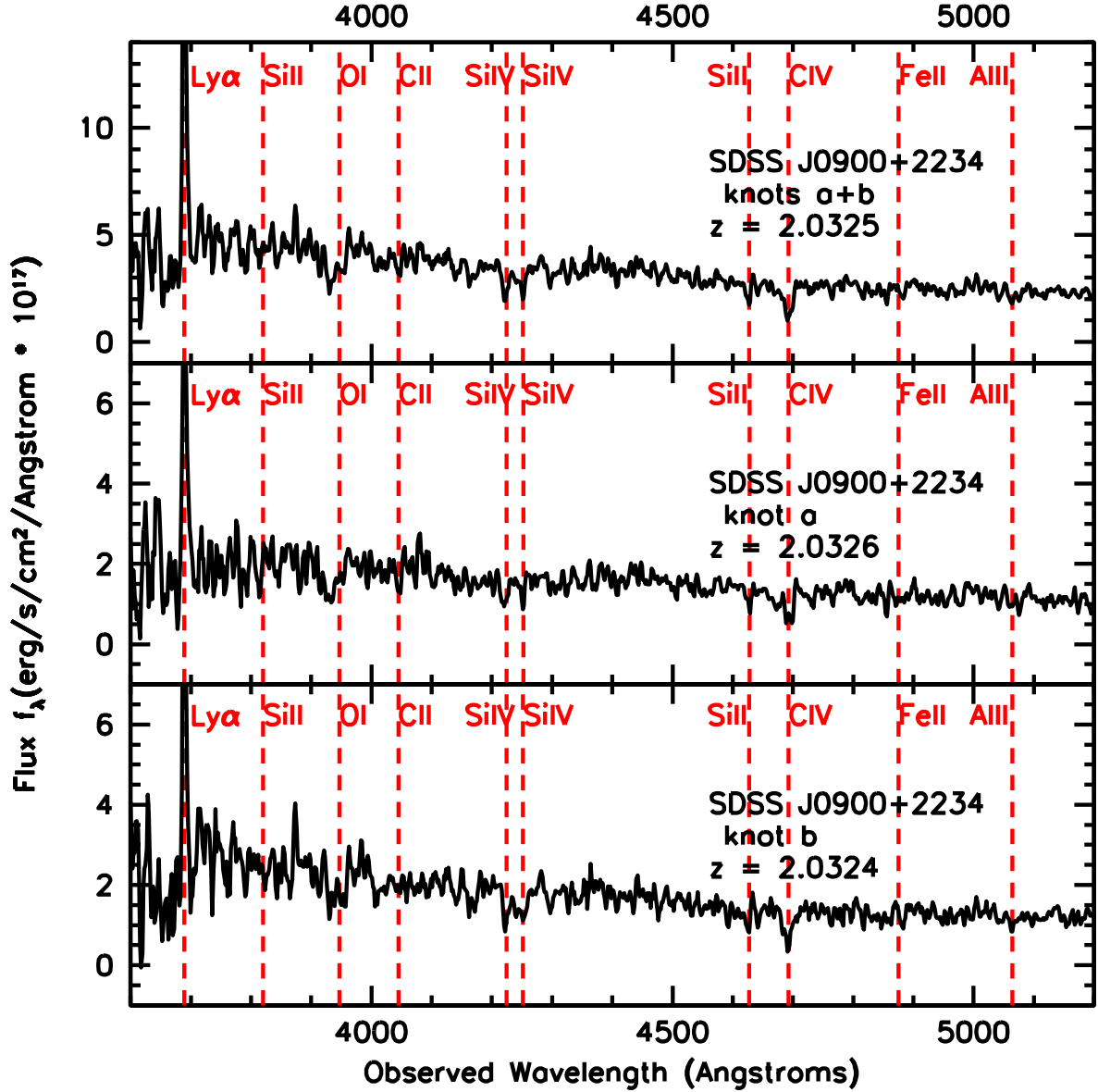}
\caption{Spectra for  SDSS J090002.79+223403.6. 
In the upper panel we show the sum of the spectra from knots a and b.
as shown in Fig. 2 (right). In the middle panel we show the spectrum
of knot a. In the lower panel we show the spectrum from knot b.
\label{fig-6}}
\end{figure}

We report results on four systems that were found to have source redshifts 
$z>2.0$. 
Figures~\ref{fig-1} and \ref{fig-2} show color-coadd images of the 
four systems as observed from the ARC 3.5m telescope using SPICAM. 
Figures~\ref{fig-3} -- \ref{fig-6} show the spectra of the source arcs 
from the ARC 3.5m using DIS III.  Table~\ref{tab-sys} provides the 
positions, redshifts and magnitudes of lensed images of the source 
galaxy.  The Einstein radius ($\Theta_E$) is estimated from a simple manual fit
of a chord of a circle to the arc. Using $\Theta_E$ and the spectroscopically 
measured source and lens redshifts, we can estimate the mass 
enclosed within the Einstein radius ($M(<\Theta_E)$). 
Here, a singular isothermal sphere (SIS) was used to approximate the lenses.
This model is typically used for cluster-scale masses~\cite{NB}. 
A flat cosmology with $\Omega_{\Lambda}=0.7$, $\Omega_{M}=0.3$,
     and $H_0 = 70$ km/s/Mpc $(h=0.7$) is assumed for these calculations.
These results are summarized in Table~\ref{tab-sys2}.

\subsection{SDSS J090122.37+181432.3}
SDSS J090122.37+181432.3 was selected based on the fact that it has three 
additional galaxies located within $10 \arcsec$ of the central LRG 
that have a color $g$-$r$ $< 1$ and $r$-$i$ $< 1$ $(N=3)$.
Fig.~\ref{fig-1} (left) shows the $g,r,i$ false color 
image of the system. There 
is a bright central LRG (SDSS J090122.37+181432.3)
with many galaxies in the same cluster located 
nearby. The SDSS DR7 data provides the spectroscopic redshift  
$z=0.3459\pm0.0003$ of the central LRG.  

The source is comprised of 
a long arc to the east of the central LRG. DIS spectroscopy provides 
the redshifts of the arc components to the northeast and southeast 
of the central LRG. The redshifts are consistent with the interpretation
the arc is a single system. The mean redshift of the two components 
was $z=2.2558\pm0.0003$. The uncertainty on the redshift comes from the 
full spread of the measurements in slightly different setups of the slit.
Fig.~\ref{fig-3} shows that the spectra of the two components
exhibit clear and strong emission, which we identify as Ly$\alpha$.
[OII] 3727 at low redshift is ruled out by the absence of expected
accompanying strong emission lines from [OIII] 5007 and in particular 
H$\alpha$.  The Ly$\alpha$ identification and hence the redshift are also
confirmed by clear detections of rest-frame optical emission lines 
from near-IR spectroscopic observations~\cite{hainline09}.
The figure also marks the expected positions of typical LBG absorption 
features in the rest-frame UV~\cite{shapley03}, but these lines are at best 
only weakly detected in the spectrum of knot b (maybe SiIV and CIV in the 
upper panel).

The Einstein radius of the arc, is $7.7\pm1.1\arcsec$. The uncertainty is a 
result of the change in curvature of the arc over its length.
The mass within the Einstein radius is $9.6\pm2.7\times 10^{12}$
M$_{\sun}$. 
The uncertainty in the mass is
dominated by the contribution from the uncertainty in the Einstein
radius. That is the case for all of the systems described in this paper.

\subsection{SDSS J134332.85+415503.4}
SDSS J134332.85+415503.4 comes from the merging galaxy sample.
Fig.~\ref{fig-1} (right) shows the false color image of the system, which 
is comprised of a rich galaxy cluster with a bright central LRG. 
The SDSS catalog provides 
the spectroscopic redshift of two central galaxies that are within 
$10 \arcsec$ of the lensed galaxy. The bright, central galaxy 
(SDSS J134332.85+415503.4) has redshift
$z=0.4180\pm 0.0002$. The second-brightest galaxy, located immediately
southeast of the central LRG, has redshift 
$z=0.4135\pm 0.0004$. 

The arc, approximately $20\arcsec$ in length, 
is to the east and northeast of the 
central LRG.  DIS spectroscopy of the arc 
provides $z=2.0927\pm0.0003$, where the uncertainty is based 
on the measured differences of the redshifts of knots in other systems. 
The spectrum is shown in Fig.~\ref{fig-4}.  For comparison, this figure
also shows, shifted to the same redshift, the spectrum of the composite 
LBG template \citep{shapley03} that we are using for cross correlation
and redshift measurement.  Note the clear presence in the arc spectrum of
all the marked typical LBG absorption features that are seen in the template.
However, also note that the arc spectrum shows Ly$\alpha$ absorption instead
of emission as in the template, and for this system (only) we have
limited the cross correlation wavelength range to 3800 -- 5500 \AA \
in order to avoid the Ly$\alpha$ emission line in the LBG template. 

A simple fit determined that the Einstein radius is $12.7\pm0.4 \arcsec$.
The mass within the Einstein radius is $31.7\pm 2.0\times 10^{12}$ 
M$_{\sun}$. 

\subsection{SDSS J114833.14+193003.2}
SDSS J114833.14+193003.2, a.k.a. ``The Cosmic Horseshoe''~\cite{belokurov}, 
was selected from the $N=3$ sample in our 
search around LRGs. Fig.~\ref{fig-2} (left) shows the false color image.
The SDSS DR7 provides spectroscopic redshift $z=0.4457\pm0.0003$ 
for the central bright galaxy (SDSS J114833.14+193003.2). 
We emphasize that we identified and confirmed this
    as a strong-lensing system prior to our knowledge that another
    group was studying it.

The arc forms a nearly continuous ring around a central LRG.
DIS spectroscopy provides an average redshift 
$z=2.3811\pm0.0003$ for the three bright  knots arrayed around the
central LRG. Again, the uncertainty is due to  differences
between individual redshifts of the knots. The spectra are 
shown in Fig.~\ref{fig-5}, where we see strong Ly$\alpha$ emission
and the presence of most of the marked typical LBG absorption features.

A simple fit determined that the Einstein radius is $5.0\pm0.3\arcsec$.
The mass within the Einstein radius is $5.1\pm0.6\times 10^{12}$
M$_{\sun}$. The Einstein radius and enclosed mass 
are in agreement with those previously published~\cite{belokurov}.

\subsection{SDSS J090002.79+223403.6}
SDSS J090002.79+223403.6 was selected from the $N=4$ sample in our search
around LRGs. Fig.~\ref{fig-2} (right) shows the  false color image
of the system, which has two bright central galaxies.
The SDSS data provides the spectroscopic redshift, 
$z= 0.4890\pm0.0002$, for the more northwest of those two galaxies.

The source galaxy appears as a knot to the east and an arc 
to the southeast of the lens and as a knot immediately to the west of the lens.
Analysis of the DIS spectra, shown in Fig.~\ref{fig-6},
indicates that the redshifts of the knots to the northeast and northwest 
of the central LRG pair are the same within uncertainties. The 
mean is $z=2.0325\pm0.0003$. 
Again, we take an uncertainty based on differences between 
redshifts of knots in the other systems described in this paper.
The spectra again show strong Ly$\alpha$ emission corroborated by
detection of some of the typical LBG absorption features, in particular
the pair of SiIV lines and the CIV line.

The simple fit determined that the Einstein radius is $7.0\pm 0.8 \arcsec$.
The mass within the Einstein radius is $11.6\pm 2.7\times 10^{12}$
M$_{\sun}$.

\section{Summary and Conclusion}
Our strong lens search has yielded four systems with lensed  
galaxies that have redshift $z>2.0$. The lens redshifts are presented, 
based on the SDSS spectroscopic data. We measured the source galaxy redshifts 
using the ARC 3.5m DIS spectrograph. Further imaging using the 
ARC 3.5m imager SPICAM revealed more detail about the lensed 
systems.  The arcs from the lensed galaxies have Einstein radius 
between $5.0\arcsec$ and $12.7\arcsec$ and are quite bright.
A simple SIS model was used to provide estimates 
of the mass enclosed within the Einstein radius
of the lensing galaxies.

From the data in Table~\ref{tab-sys}, we see that these systems have
approximate total isophotal magnitudes $r = 19.2 - 20.4$.
For comparison with other lensed systems, we use the
$3\arcsec$-diameter magnitudes and those are $r = 20.0 - 20.6$. 
Thus these systems are about one magnitude fainter than the
brightest lensed LBG known, the 8 o'clock arc 
($r = 19.2$)~\cite{8oclock}, 
but are quite similar to the few other very bright $z > 2$ 
lensed systems, such as MS 1512-cB58 ($r = 20.4$)~\cite{cB58} 
and the Clone ($r = 19.8$) ~\cite{clone}.  Likewise, the
brightest knots in these systems also have very high surface
    brightness, $\sim 23.5 \; {\rm mag \; arcsec}^{-2}$ in $r$, similar to
    the Clone, and about one magnitude fainter than the 8 o'clock arc.  
    The systems in our sample are thus some of the brightest lensed
    galaxies known at $z > 2$ and will be amenable to detailed 
    studies of their individual properties.

Combined with our previous results~\cite{8oclock, clone, kubo}
the Sloan Bright Arcs Survey has now 
spectroscopically confirmed  12 strong lensing systems in SDSS DR5 and DR6.
There are follow-up observations of all of these systems scheduled 
for either Cycle 16 Supplemental data or 
for Cycle 17 of the Hubble Space Telescope. These observations
will provide high-quality images useful for detailed lens and source 
modeling. Additional follow-up observations obtained using WIYN 3.5m
telescope at Kitt Peak National Observatory may provide wide-field images 
of the lensing galaxy that can be used to characterize the galaxy
cluster environment.  We continue to follow up additional promising
lens candidates.

\acknowledgments


Fermilab is operated by the Fermi Research Alliance, LLC under Contract No.
DE-AC02-07CH11359 with the United States Department of Energy. These results
are based on observations obtained with the SDSS and the Apache Point
Observatory 3.5-m telescope, which is owned and operated by the 
Astrophysical Research Consortium. Funding for the SDSS and SDSS-II has been
provided by the Alfred P. Sloan Foundation, the Participating Institutions,
the National Science Foundation, the U.S. Dept. of Energy, the National 
Aeronautics and Space Administration, the Japanese Monbukagakusho, the 
Max Planck Society, and the Higher Education Funding Council for England. 
The SDSS Web Site is http://www.sdss.org/.






\clearpage

\clearpage

\clearpage

\clearpage



\begin{table}
\tablewidth{0pt}
\caption{\label{tab-obs}
Imaging observation Log. Units of right ascension and declination  
are degrees.}
\begin{tabular}{|c|lcclc|} \hline
System               & Date & Instrument 
                             & Filter         & Exposure      & Seeing \\
RA and DEC (J2000)  
                     &      &&                &  Time         & $g,r,i$ \\ \tableline
SDSS J090122.37+181432.3      
                     & 2007 Apr 21     
                            & SPICAM 
                             & $g,r,i$        & $3\times 300$ 
                              & $1.0\arcsec , 0.8\arcsec , 0.8\arcsec$ \\ 
135.3432     18.2423 &      &&                &               & \\ \tableline
SDSS J134332.85+415503.4 
                      & 2008 Jun 9 
                            & SPICAM 
                             & $g,i$          & $2\times 300$ 
                              & $1.4\arcsec , 1.2\arcsec , 1.4\arcsec$ \\ 
205.8869    41.9176   & 2008 Jun 11 
                            & SPICAM 
                             & $g,r$          & $1,3\times 300$ & 
                                                      \\ \tableline
SDSS J114833.14+193003.2 
                     & 2008 Jan 11 
                            & SPICAM          
                             & $g,r,i$        & $3\times 300$ 
                              & $1.2\arcsec , 1.0\arcsec , 0.9\arcsec$ \\ 
177.1381    19.5008  &      &&                &  &     \\ \tableline
SDSS J090002.79+223403.6 
                     & 2006 Nov 22 
                            & SPICAM 
                             & $g,r,i$        & $3\times 300$
                              & $0.9\arcsec , 1.0\arcsec , 1.3\arcsec$ \\ 
135.0116    22.5677  &      &&                &   &    \\  \tableline
\end{tabular}
\end{table}

\begin{table}
\tablewidth{0pt}
\caption{\label{tab-spe}
Spectral observation Log. }
\begin{tabular}{|c|lccl|} \hline
System               & Date & Instrument 
                             & Grating         & Exposure Time \\ \tableline 
SDSS J090122.37+181432.3      
                     & 2007 Mar 10     
                            & DIS 
                             & R400/B300      & $3\times 900$ \\       
                     & 2008 Mar 11     
                            & DIS 
                             & R400/B300      & $1\times 900$ \\   \tableline
SDSS J134332.85+415503.4 
                     & 2008 Mar 11 
                            & DIS 
                             & R400/B300      & $2\times 900$  \\ 
                     &      &&                & $1\times 740$  \\ \tableline
SDSS J114833.14+193003.2 
                     & 2007 Mar 10 
                            & DIS          
                             & R400/B300      & $2\times 900$ 
                                                       \\ \tableline
SDSS J090002.79+223403.6 
                     & 2007 Mar 10 
                            & DIS 
                             & R400/B300       & $3\times 900$ 
                                                       \\ \tableline
\end{tabular}
\end{table}

\begin{table}
\tablewidth{0pt}
\caption{Strong-lens system properties including the label, right ascension
and declination (J2000), and the magnitudes (both isophotal and 
$3 \arcsec$-diameter circular aperture) 
of the segments of the source arcs in the $g$, $r$, and $i$-band filters. 
Note that for SDSS J134332.85+415503.4, instead of a $3 \arcsec$-diameter
circular aperture, we used a curved aperture 
(a partial elliptical annulus with width $2.5 \arcsec$
and length $21.8 \arcsec$) that follows the arc.  The ``total'' magnitudes
reported are simply the sum of the magnitudes
    for the listed arc components.  The $r$-band isophotal limits (mag. per
    arc-seconds$^2$) are, in order, 25.0, 25.4, 25.5 and 25.2. 
The statistical uncertainties in the magnitudes are $< 0.05$ mag.
    The units of right ascension and 
    declination  are degrees.
 \label{tab-sys} }
\begin{tabular} {|c|c|c|c|ccc|ccc|} \hline
       & Arc  & &     & \multicolumn{3}{c|}{$3\arcsec$ mag.}
                                           &\multicolumn{3}{c|}
                                            {isophotal mag.}  \\ 
SDSS System 
       & Segment 
              & RA  
                & DEC & $g$  & $r$  & $i$  & $g$ & $r$ & $i$  \\ \hline
J090122.37+181432.3                          
       & b    & 135.344265 
                & 18.240161     
                      & 22.0 & 21.1 & 20.6 & 21.4&20.4 &20.0  \\ 
       & d    & 135.344531 
                & 18.243560     
                      & 22.5 & 21.8 & 21.3 &21.5 &20.6 &20.1  \\ 
       & Total& &     & 21.5 & 20.6 & 20.1 &20.7 &19.7 &19.3  \\ \hline
J134332.85+415503.4                          
       & NA     & 
                &     & 20.6 & 20.1 & 19.9 &20.9 &20.4 &20.2  \\ \hline
J114833.14+193003.2                          
       & a    &  177.138643
                & 19.499833
                      & 21.9 & 21.3 & 21.0 &21.3 &20.7 &20.3   \\ 
       & b+d  &  177.136772
                & 19.501084
                      & 21.9 & 21.3 & 21.0 &20.5 &19.9 &19.6   \\ 
       & c    &  177.138971
                & 19.502187
                      & 22.3 & 21.7 & 21.5 &21.6 &21.0 &20.7   \\ 
       & Total& &     & 20.8 & 20.2 & 19.9 &19.8 &19.2 &18.9   \\ \hline
J090002.79+223403.6                          
       & a    & 135.013829   
                &  22.568782      
                      & 21.2 & 21.2 & 21.2 &21.0 &20.9 &20.9   \\ 
       & b    & 135.009554   
                &  22.568546      
                      & 21.5 & 21.3 & 21.2 &21.3 &20.8 &20.5   \\
       & c    & 135.011499   
                &  22.565599    
                      & 22.3 & 21.6 & 21.3 &21.9 &21.3 &21.0   \\
       & d    & 135.012918    
                &  22.566221      
                      & 22.7 & 22.4 & 22.2 &22.3 &21.9 &21.7   \\ 
       & Total& &     & 20.3 & 20.0 & 19.9 &20.0 &19.6 &19.4   \\ \hline
\end{tabular}
\end{table}

\begin{table}
\tablewidth{0pt}
\caption{Strong-lens system properties including average redshift of the
source, the Einstein radius, and enclosed mass of the lens.  As noted in
the text, we used $h=0.7$. 
\label{tab-sys2} }
\begin{tabular}{lcccc} \hline
Lens/Source
                    &$z_{source}$ &$z_{lens}$
                       & $\Theta_E$ & $M(<\Theta_E)$ \\ 
                    &  & & arcsec
                                    & $\times 10^{12}$ M$_{\sun}$
                                                           \\ \hline
SDSS J090122.37+181432.3 
                    &$2.2558$ & $0.3459$ 
                       &$7.7\pm1.1$ 
                                    &$9.6\pm2.7$   \\
SDSS J134332.85+415503.4 
                    &$2.0927$ &$0.4135$ 
                       &$12.7\pm0.4$        
                                    &$31.7\pm2.0$   \\
SDSS J114833.14+193003.2 
                     &$2.3811$ &$0.4457$
                       &$5.0\pm0.3$
                                    &$5.1\pm0.6$    \\
SDSS J090002.79+223403.6 
                     &$2.0325$ & $0.4890$ 
                       &$7.0\pm0.8$ 
                                    &$11.6\pm2.7$   \\ \hline
\end{tabular}
\end{table}




\end{document}